# How does the connectivity of open-framework conglomerates within multi-scale hierarchical fluvial architecture affect oil sweep efficiency in waterflooding?

Naum I. Gershenzon[1], Mohamad Reza Soltanian[1], Robert W. Ritzi Jr.[1], David F. Dominic[1], Don Keefer[2], Eric Shaffer[2], Brynne Storsved[2]

[1] *Department of Earth and Environmental Sciences, Wright State University, 3640 Col. Glenn Hwy., Dayton, OH 45435*

[2]*Illinois State Geological Survey, University of Illinois, Urbana-Champaign*

**ABSTRACT**

We studied the effects on oil sweep efficiency of the proportion, hierarchical organization, and connectivity of high-permeability open-framework conglomerate (OFC) cross-sets within the multi-scale stratal architecture found in fluvial deposits. Utilizing numerical simulations and the RVA/Paraview open-source visualization package, we analyzed oil production rate, water breakthrough time, and spatial and temporal distribution of residual oil saturation. The effective permeability of the reservoir exhibits large-scale anisotropy created by the organization of OFC cross-sets within unit bars, and the organization of unit bars within compound bars. As a result oil sweep efficiency critically depends on the direction of the pressure gradient. When pressure gradient is oriented normal to paleoflow direction, the total oil production and the water breakthrough time are larger, and remaining oil saturation is smaller. This result is found regardless of the proportion or connectivity of the OFC cross-sets, within the ranges examined. Contrary to expectations, the total amount of trapped oil due to the effect of capillary trapping does not depend on the pressure gradient within the examined range. Hence the pressure difference between production and injection wells does not affect sweep efficiency, although the spatial distribution of oil remaining in the reservoir depends on this value. Whether or not clusters of connected OFC span the domain does not affect sweep efficiency, only the absolute rate of oil production. The RVA/Paraview application allowed us to visualize and examine these non-intuitive results.

# INTRODUCTION

Oil sweep efficiency is fundamentally controlled by the nature of immiscible displacement of a non-wetting liquid by a wetting liquid in porous media (Buckley and Leverett, 1942). This process includes the effects of capillary pressure and relative permeability on oil trapping and early water breakthrough (Kortekaas, 1985; Corbett et al., 1992; Khataniar and Peters, 1992; Wu et al., 1993; Gharbi et al., 1997; Kaasschieter, 1999). Such immiscible displacement is, in turn, controlled by the three-dimensional sedimentary architecture of reservoirs, which creates heterogeneity and anisotropy in permeability (Kjonsvik et al., 1994; Jones et al., 1995; Tye et al., 2003; Choi et al., 2011).

Recent studies have led to new conceptual and quantitative models for sedimentary architecture in fluvial deposits (Fig.1) over a range of scales that are relevant to the performance of some important petroleum reservoirs (Tye et al., 2003; Lunt et al., 2004; Bridge, 2006; Lunt and Bridge 2007). Based on the new understanding of hierarchical and multi-scaled sedimentary architecture, we created digital realizations of the three-dimensional (3D) fluvial-type reservoir and investigated how the spatial variations in reservoir properties (relative permeability and capillary pressure) affect oil sweep efficiency in waterflooding (Gershenzon et al., 2014). Analysis of the simulation results reveals nuances complementary and sometimes contradictory to some of the conventional understanding of oil recovery by waterflooding.

Indeed, previously it was shown that during immiscible oil displacement, oil may be trapped due to reservoir heterogeneities in permeability and capillary pressure (Kortekaas, 1985; Wu et al., 1993; Kaasschieter, 1999). The trapping effect, hence oil sweep efficiency, depends on the pressure gradient (Corbett et al., 1992). However, we found that the value of the pressure gradient does not practically affect oil sweep efficiency although the spatial distribution of oil remaining in the reservoir depends on this value (Gershenzon et al., 2014). In the reservoir models we considered (Gershenzon et al., 2014), capillary pressure differs between sandstone with high capillary pressure and open-framework conglomerate (OFC) with low capillary pressure, such that residual oil saturation is always lower in sandstone. The interplay of processes controlled by the interconnection of units of OFC and those resulting from the juxtaposition of sandstone and OFC defines the spatial and temporal distribution of oil. Given the heterogeneity within these models, it is surprising to find that oil saturation averaged over the reservoir does not depend on pressure difference, in contrast to what was known previously (i.e., Corbett et al., 1992). In essence, the effect of oil trapping in isolated OFC clusters is canceled by the effect of oil moving out of surrounding sandstone and through connected OFC clusters.

The hierarchical stratal architecture causes the connectivity of higher-permeability OFC cells to differ with scale and direction, and creates anisotropy in the bulk effective permeability. Since the difference in permeability between OFC and sandstone is up to four orders of magnitude and since the size of OFC clusters considerably exceeds the cell size, it is expected that the water-oil boundary would show large-scale fingering. However, we did not observe fingering in all tested reservoir realizations. Even so, most of the oil (80-95%) reaches the production well through the OFC clusters (Gershenzon et al., 2014).

The results of waterflooding simulations with pressure gradient aligned parallel to the paleoflow direction are considerably different from those with pressure gradient aligned normal to paleoflow direction in oil sweep efficiency. Thus, while the hierarchical organization of OFC cross strata within larger-scale deposits does not promote large-scale fingering, it does impart significant anisotropy. We found that the volume of water injected before water breakthrough is considerably larger (by a factor 1.5) when the pressure gradient is in the normal to paleoflow direction than when the gradient is in the paleoflow direction, regardless of the pressure difference (Gershenzon et al., 2014). While anisotropy in fluvial reservoirs is expected the anisotropy effect on waterflooding is not entirely understood.

Finally, we found that oil sweep efficiency does not practically depend on the proportion of open-framework conglomerate in the reservoir. When the proportion of OFC is lower than about 20%, clusters of connected OFC cells do not span the reservoir. Even so, the absence of spanning clusters of OFC does not abruptly affect oil sweep efficiency.

The above-mentioned specific properties of waterflooding processes are not intuitive and need to be further explored. The goal of this article is to further investigate these results and to provide a more complete description of the observed processes. The Reservoir Visualization Analysis (RVA) package (Keefer et al., 2012) built on the ParaView open-source platform was used to further visualize, analyze, and illustrate our results. In this article, we first provide a more complete description of the reservoir model, then present the methodology for reservoir simulation and visualization of the results, followed by results, discussion and conclusions.

**RESERVOIR MODEL**

The sedimentary architecture of reservoirs created by fluvial deposition has recently been described in three-dimensional models (Bridge, 2006). As shown in Fig. 1, channel-belt deposits are characterized by a large volume fraction of convex-up, bar deposits formed within channels. Only when channels are abandoned and filled under lower flows are "channel-shaped" units formed. These concave-up channel fills are low-permeability baffles within the domain. In gravelly channel-belt deposits, preferential flow pathways arise from the interconnection of open-framework gravels (the unlithified equivalent of open-framework conglomerates or OFC) within lobate unit bar deposits (Lunt and Bridge, 2007). These are the "thief zones" that have a negative effect on oil recovery (McGuire et al., 1994; 1999; Tye et al., 2003).

Lunt et al. (2004) studied the gravelly channel belt of the Sagavanirktok River (Alaska, Fig. 1), a modern analog for deposition of the Ivishak Formation in Prudhoe Bay, and quantified the proportions and lengths for unit types across relevant scales (Table 1). At the smallest scale, sets of cross-stratified ("cross-sets" herein) sand, sandy gravel, and open-framework gravel (decimeters thick and meters long) occur within unit bar deposits (tens of decimeters thick and tens of meters long). Unit bars and cross-bar channel fills occur within compound bar deposits

(meters thick and hundreds of meters long). Compound bar deposits and the channel fills that bound them occur within channel belts (tens of meters thick and kilometers long). Importantly, the open-framework gravel cross-sets were found to make up 25 to 30 percent of the volume of the deposit. These sedimentary unit types are preserved with similar lengths and proportions within the Victor interval of the Ivishak Formation. (Tye et al., 2003). The Victor interval has few shales or other vertical permeability barriers, and connected cross-sets of open-framework conglomerates (lithified open-framework gravels) are the dominant control on reservoir performance (McGuire et al., 1994, 1999; Tye et al., 2003).

The sedimentary architecture quantified by Lunt et al. (2004) was incorporated by Ramanathan et al. (2010) into a high-resolution, three-dimensional, digital model using geometric-based simulation methods. Guin et al. (2010) confirmed that the model contains the hierarchy of sedimentary unit types and honors the proportions, geometries, and spatial distribution of the unit types quantified at each level by Lunt et al. (2004).

In this model, cross-sets of high-permeability open-framework conglomerate (OFC) were simulated discretely. OFC cells are considered to be connected when cell faces are adjacent. Importantly, clusters of continuously connected OFC cells create preferential flow pathways. When OFC cross-sets comprise at least 20% of the volume of the deposit, clusters span opposing pairs of domain boundaries (Guin et al., 2010). Such spanning, preferential-flow pathways have been inferred to exist within the Ivishak Formation (Tye et al., 2003).

The number, size, and orientation of OFC clusters in the model change with proportion. At any given proportion, the number, size, and orientation of clusters change across the different hierarchical levels (scales) of the stratal architecture. Connected OFC cells within individual cross-sets form paths that vertically span single unit bar deposits. Connections across unit bar boundaries enhance lateral branching and the many clusters within unit bars connect into a smaller number of larger clusters at the scale of multiple unit bars. At the scale of an entire compound bar deposit, these clusters are typically connected into one or two large, spanning clusters. The spanning clusters occur at proportions of OFC cells below the theoretical threshold (31%) for random infinite media predicted in the mathematical theory of percolation. Guin and Ritzi (2008) showed that this is caused by geological structure within a finite domain. The percolation theory shows why two-dimensional models under-represent true three-dimensional connectivity, and thus why simulations must be three-dimensional (Huang et al., 2012).

The saturated permeability in sandy-gravel deposits (and of their lithified equivalents, pebbly sandstones and sandy conglomerates) varies non-linearly as a function of the volume of sand mixed with gravel (see Fig. 6 in Ramanathan et al., 2010; Klingbeil et al., 1999; Conrad et al., 2008). Sandy-gravel strata have permeabilities similar to the sand they contain, which are of the order of $10^0$ to $10^1$ Darcies. Thus, sand and sandy-gravel cross-sets within unit bars have permeabilities similar to channel-fill sands. Open-framework gravels have permeabilities of the order of $10^3$ to $10^4$ Darcies. In either type of strata, the coefficient of variation in permeability is of the order of unity. In the lithified stratatypes within the Ivishak Formation (sandstones, pebbly sandstones and open-framework conglomerates), the saturated permeabilities scale down accordingly (Tye et al., 2003). In this study, the geometric mean for saturated permeability of sandstones and pebbly sandstones is taken to be 66 milli-Darcies (mD). Because these two lithotypes are assigned the same permeability distribution in the simulations, we refer to them collectively as "sandstone." The geometric mean saturated permeability of open-framework conglomerates is taken to be 5250 mD. Within all these lithotypes, the permeability distributions were created with a coefficient of variation of unity as is found in natural deposits.

**METHODOLOGY**

Using a code by Ramanathan et al. (2010) we generated six realizations of a reservoir volume, each with a different proportion of OFC (Table 2). As shown in Table 2, the proportion of OFC controls the mean permeability, and also affects the proportion of high-permeability cells that are connected to create preferential flow pathways. We consider clusters to span opposing boundaries when their extent in $x$, $y$ and $z$ directions is equal to the reservoir size in those directions. Thus, the reservoir models range from one with more than 90% of all OFC cells connected in one spanning cluster (realization 1) to those with no spanning cluster (realizations 5 and 6). Fig. 2 shows a 3D view of realizations 1, 4 and 6 with 28%, 22% and 16% of OFC cells, respectively. The blue color indicates the OFC cells. Although it is not obvious from the Figs., realization 1 (bottom panel) incorporates almost all OFC cells (90%) in one spanning cluster, realization 4 (middle panel) has one spanning cluster (55% of all OFC cells) and realization 6 (top) has no spanning clusters. Fig. 3A displays the same realizations but only OFC cells are shown. Small parts (1/64) of these reservoirs are shown on Fig. 3B. These Figs illustrate that OFC cells are densely interconnected in all considered cases, although the distributions of OFC material between spanning and non-spanning clusters are very different among these realizations. As we will demonstrate later, the existence of this "small-scale dense net" explains the main features of waterflooding in fluvial reservoirs. Fig. 4 depicts the distribution of non-spanning clusters by size for these three realizations. The distribution exhibits similarity among these cases, i.e. the proportion of nonspanning clusters of any size is almost identical. Fig. 5 shows the five largest clusters of realization 4; the largest cluster is a spanning cluster. Fig. 6 demonstrates permeability distribution along with lithotype distribution.

Utilizing these realizations, we simulated 3D immiscible oil displacement by water (black oil approximation) with ECLIPSE (Schlumberger Reservoir simulation software, version 2010.2). The approach was as follows. The top of the parallelepiped reservoir model is taken to be at a depth of 2560 m (8400 feet). The size of the reservoir in the $x$, $y$ and $z$ directions are $L_x$=200 m (656 feet), $L_y$=200 m (656 feet) and $L_z$=5 m (16.4 feet), respectively. This represents the heterogeneity created by an assemblage of unit bars within a compound bar. The highly non-linear nature of the computational problem currently limits us to simulations of this reservoir size. Our goal is to eventually grow the problem to larger domains with finer resolution. Initially the reservoir contains oil with dissolved gas and connate water in equilibrium. The reservoir is divided into one million cells with cell size 2 m x 2 m x 0.05 m in the $x$, $y$ and $z$ directions, respectively. The permeability of each cell is assigned from the appropriate distribution defined for each lithotype in the previous section. Fig. 6 shows 3D views of permeability distribution along with views of the lithotype distribution. The distribution of permeability values is illustrated in Figs.7 and 8 for the realizations 1 and 6, respectively. The percentage of cells with permeability in each specified range is also shown. While the percentage of high-permeable material is different between these two cases, the common features are: (1) The majority of cells (69.2% and 80.2%) have permeability less than 1 D (the geometrically averaged permeability for the realizations 6 and 1 are 149 and 226 mD). (2) The second maximum in permeability distribution is in the range from 3 to 4 D. The latter range brackets the geometric average of permeability of all OFC cells. (3) The cells with permeability inside of each range connect to small-scale (compared with reservoir size) nets in all ten considered ranges of permeability.

Two different property tables were utilized for the sand and OFC lithotypes. These tables define the relations between water relative permeability, oil-in-water relative permeability, water-oil capillary pressure, gas relative permeability, and water saturation. Figs. 5 and 6 (from Gershenzon et al.,2014) show, respectively, the relative permeability and capillary pressure as a function of water saturation for sand and OFC lithotypes. (See Gershenzon et al., 2014 for more details.)

## RESULTS AND DISCUSSION

### Anisotropy

Fluvial architecture creates anisotropy in which the bulk effective permeability is different in the paleoflow direction ($y$) than in the direction normal to the paleoflow ($x$) (Guin et al., 2010). Also, the effective permeability in the horizontal direction is very different from that in the vertical direction ($z$). To consider how anisotropy affects the process of oil displacement, we compared results of simulations with the pressure gradient along the $y$ direction to the same case with the pressure gradient along the $x$ direction. Gershenzon et al. (2014.) showed that the results are essentially the same for gradients in the positive and the negative coordinate directions in each case. Therefore, the dip direction of the OFC strata is not a factor, and we only show results for the positive directions. Fig. 9 clearly illustrates the effects of anisotropy. Under the same pressure gradient between injection and production wells, the waterflood front propagates faster when the pressure gradient is in the $y$ direction. The front is broader when the pressure gradient is in the $x$ direction, so that a larger volume of the reservoir is swept (see animated Fig.10). This qualitative result illustrates why sweep efficiency increases if the injector and producer are aligned normal to the paleoflow direction, and supports the values for sweep efficiency reported in Gershenzon et al. (2014; see their Figs. 7-10 and Table 3).

Anisotropy can be understood in terms of the sizes of the sandstone and OFC clusters in $x$, $y$ and $z$ directions (see Table 2). The mean size of both sandstone and OFC clusters in the paleoflow direction is almost double the size of the clusters normal to paleoflow horizontal direction for all realizations. Figs. 5, 11 and 12 clearly illustrate that most of the clusters are elongated the in paleoflow direction. Anisotropy is caused by the way smaller-scale OFC strata are organized within the stratal architecture at larger scales.

Anisotropy in the vertical direction is even more dramatic. The difference in cluster size between horizontal and vertical directions ranges from 20 to 40 times (see Table 2, note that we compare size in length units not in cell number units).

**Connectivity**

The existence and connectivity (clusterization) of high-permeability material in a fluvial reservoir plays a major role in waterflooding processes. Indeed, as we showed previously, from 80% to 95% of oil comes to a production well through the OFC cells even though the total proportion of OFC cells is from 16% to 28%. This is not a surprise since the permeability of OFC is from two to four orders of magnitude larger than in sandstone. This difference in permeability could be expected to produce: (1) highly developed fingering of the water-oil front and (2) the pronounced difference in dynamics between reservoir realizations which include spanning OFC clusters and those without such clusters. However, Gershenzon et al. (2014; their figs. 9 and 11) showed that the oil production dynamics are not noticeably different between realizations with spanning OFC clusters (realizations 1-4) and those without spanning clusters (realizations 5 and 6). Moreover, the large-scale fingering of the water-oil front was not observed in any of the realizations. Indeed, as Fig. 13 shows, the water-oil fronts have a relatively smooth shape and are not much different between realization 1 (90 % of OFC cells in one spanning cluster) and realization 6 (no spanning cluster), regardless of the value of the pressure gradient. These features can be explained by the specific structure of the fluvial-type reservoir as follows. Note that in all reservoir realizations (1) small distances separate OFC clusters (and different branches of the same cluster) in the vertical direction (from 35 cm to 55 cm, see Table 2, column 7); and (2) the branches of OFC clusters are thin in the vertical direction (~10–15 cm). Fig.14 illustrates the typical pattern of oil flow between such structures. Oil moves preferably along the pressure gradient in high-permeability OFC clusters and diffuses preferably in the direction normal to OFC branches in the low-permeability sandstone. Since the distances between OFC branches are small and since capillary pressure pushes oil from the sandstone to the OFC, oil from sandstone cells reaches the OFC cells relatively quickly (the pressure difference between injection and production wells ranged from 100 to 800 psi). This process and the small thickness of the OFC branches explain both the absence of the large-scale fingering and the fact that most of oil reaches the production well through the OFC clusters. As Fig. 4 shows, even realizations with a small proportion of OFC material include large-scale clusters (although there are no spanning clusters). The same process of oil movement (Fig. 14) occurs in those realizations, which explains the almost identical behavior, as characterized by sweep efficiency, of waterfloding for all these reservoir realizations.

**Oil trapping**

The dynamics of immiscible oil displacement by water is described by the Buckley-Leverett equation (Buckley and Leverett, 1942). Solutions of this equation for homogeneous media and neglecting capillary pressure indicate that part of the oil (above irreducible oil saturation) remains behind the water-oil front. This is a fundamental origin of low oil sweep efficiency during the process of immiscible oil displacement. Solutions of the Buckley-Leverett equation in heterogeneous media (also neglecting capillary pressure) show that heterogeneities may "trap" oil (Wu et al., 1993; Kaasschieter, 1999), which is an additional reason for the low oil sweep efficiency. Analysis of the solutions in media with uniform permeability but heterogeneous capillary pressure revealed yet another mechanism of oil trapping (Kortekaas, 1985). The heterogeneities of various scales typical for the fluvial-type reservoirs add other complications to the process (Kjonsvik et al., 1994; Tye et al., 2003; Choi et al., 2011). The interplay of the different mechanisms makes oil sweep efficiency a complicated nonlinear function of the magnitude and direction of the pressure gradient (or velocity of water-oil front) as well as the size, direction and connectivity of heterogeneities in capillary pressure and permeability. Indeed, as our simulations show, the spatial and temporal distribution of oil saturation in fluvial type reservoirs critically depends on those parameters.

Corbett et al. (1992) modeled waterflooding in heterogeneous 2D reservoirs. They found that sweep efficiency in such reservoirs depends on the pressure gradient. The amount of oil recovered was different by a factor of two for different values of the pressure gradient. This effect was due to the difference in the capillary pressure in different materials leading to oil trapping in material with smaller capillary pressure. The amount of oil locally trapped depends on the pressure gradient ($\Delta P$) and the size of heterogeneity in the gradient direction ($l$). If the difference in capillary pressure between two materials is small compared to $\Delta P*l$, the effect of capillary pressure is also small. In contrast, if this difference is larger than $\Delta P*l$, the trapping effect is dominant. Therefore, in our simulations, the trapping effect is larger for the smallest OFC clusters and there is no trapped oil in infinite OFC clusters (see Gershenzon et al., 2014; their Figs. 12 and 13). The opposite effect also occurs when oil from the material with large capillary pressure (sandstone in our case) is pushed by the capillary pressure into the material with low capillary pressure (OFC). As a result, oil saturation in sandstone is reduced. The interplay between these two effects defines the spatial and temporal distribution of oil in a process of oil displacement by water. Thus, oil sweep efficiency is expected to depend on the pressure difference between injection and production wells, as has been concluded by Kortekaas (1985), Corbett et al. (1992), and Ringrose et al. (1993).

However, our analysis of simulations utilizing all six realizations revealed an unexpected result. In spite of the differences in the spatial distribution of oil for different pressure gradients, the integral parameters such as remaining oil saturation (hence oil sweep efficiency) did not depend on the pressure gradient (see Fig. 15). This was true regardless of the direction of the pressure gradient (i.e., parallel or normal to paleoflow direction). The effect of oil trapping in

OFC clusters is canceled by the effect of oil reduction from the surrounding sandstone. This effect occurred regardless of whether the OFC material formed finite or infinite clusters and even regardless of whether or not an infinite cluster occurred in the reservoir. Figs. 16 and 17 illustrate this result and support this conclusion. These Figs. show all cells within the reservoir having oil saturation in either of two different ranges at the same stage of the waterflooding process. Note that all cells with oil saturation in the range from 0.78 to 0.9 are the OFC cells since the irreducible saturation of the wetting phase (water) for the sandstone and OFC are different, at 0.22 and 0.1 respectively. These Figs. show that the total amount of OFC cells having such saturation of trapped oil is visibly different for the different pressure gradients for both realizations 1 and 6 (see top panels at Figs. 16 and 17). These Figs. show that the amount of trapped oil in OFC clusters is larger if the pressure gradient is smaller. In contrast, the amount of oil remaining in sandstone cells is smaller when the pressure gradient is smaller (bottom panels at Figs. 16 and 17).

**CONCLUSIONS**

The fluvial architecture represented here is not found in all reservoirs, but is found in some very important ones (e.g. the Ivishak Formation, Prudhoe Bay Field, Alaska, see, Tye et al., 2003). The basic model for hierarchical fluvial architecture used here (Lunt and Bridge, 2004; Bridge 2006) is fairly general and applies over a broad range of median grain size. Furthermore, the dimensions of unit types across the stratal hierarchy scale together with formative channel width (Bridge, 2006). An important characteristic of coarser-grained reservoirs is that open-framework conglomerate strata are connected and impart higher-permeability pathways within the reservoir (Tye et al., 2003; Guin et al., 2010). These pathways cannot be imaged in 2-D cross sections and their effect on the flow field cannot be understood from 2-D simulations (see Huang et al., 2012). To visualize and fully understand the impact of these connected flow pathways within a reservoir simulation requires advanced 3-D visualization tools such as the RVA/Paraview application used here.

Several features distinguish our approach to reservoir simulation from others:

1) The modeled heterogeneity structure and scales, hence permeability distribution, realistically reflects the typical fluvial-type reservoirs; permeability ranges over four orders of magnitude.

2) The size of reservoir heterogeneities ranges from a few cm to dozens of meters.

3) The reservoir contains two different materials: sandstone and open-framework conglomerate. These materials have different properties, requiring that two sets of property tables for simulation.

4) Capillary pressure and hysteresis effects are utilized in the simulations.

The results demonstrate that small and medium scale inclusions of high-permeability cross strata fundamentally control oil trapping processes and hence the oil sweep efficiency. The OFC material forms clusters of different sizes but with a larger extent in the paleoflow direction. This anisotropy affects the integral parameters of waterflooding such as water breakthrough time, cumulative oil production and oil sweep efficiency. Fingering on the water-oil boundary is expected because the difference in permeability between OFC and sandstone is up to four orders of magnitude, the size of some of the OFC clusters may considerably exceed the cell size, and the majority of oil reaches the production wells through OFC channels. However, large-scale fingering was not observed. We did observe different degrees of water-oil front diffusivity for the different reservoir realizations and different pressure gradients, which could be considered as small-scale fingering. Large-scale fingering is absent because the width of the sandstone layers between OFC channels is small, so there is enough time for oil to diffuse from the sandstone to OFC channels, unless the pressure gradient is extremely large.

The content, structure and connectivity of high permeability material in a fluvial-type reservoir affects both oil production rate and water breakthrough time. However, oil sweep efficiency is not practically affected by the proportion and degree of connectivity of high-permeability material. The most surprising finding of our 3D simulations is the independence of oil sweep efficiency from the pressure gradient in all considered realizations and directions of the pressure gradient. This is even more surprising if we keep in mind the highly nonlinear character of waterflooding processes. Somehow the amount of oil trapped in OFC, which decrease with increase of pressure gradient, is compensated by the amount of remaining oil in sandstone, which increase with increase of pressure gradient, in such a way that the total amount of remaining oil does not depend on the pressure gradient. This "self-regulating process" could be understood in terms of small-scale and dense networks of OFC cells. Thus, oil sweep efficiency is not affected by the pressure gradient although the spatial distribution of remaining in reservoir oil critically depends on this value.

## ACKNOWLEDGMENTS

We thank Schlumberger Company for donation of ECLIPSE Reservoir Simulation Software. The RVA visualization software was developed through support from the US Department of Energy National Energy Technology Laboratory under contract DE-FE0005961.

# Figures

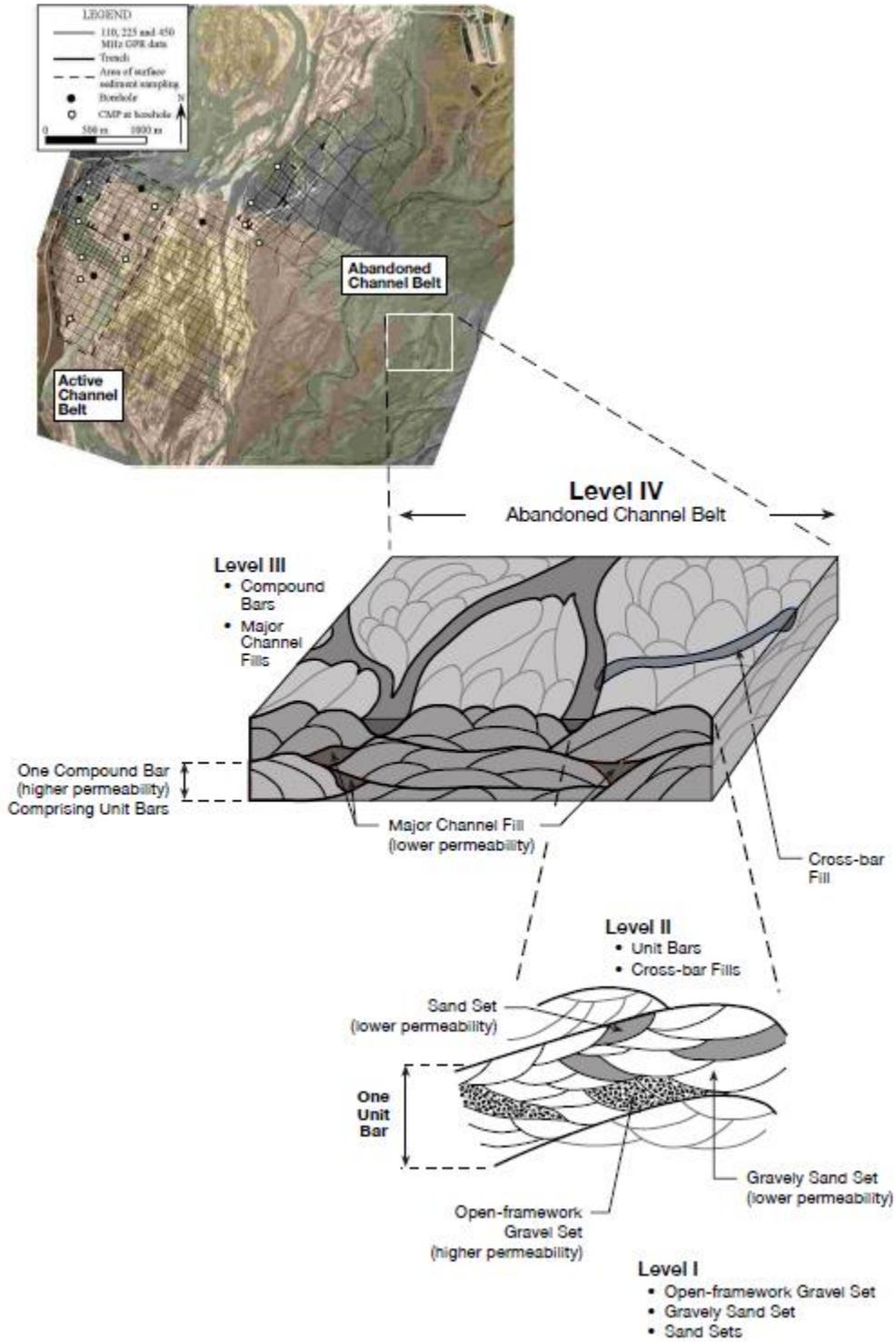

**Figure 1**. (Top) Study areas in the active channel belt and in the preserved channel belt deposits of the Sagavanirktok River (Lunt et al., 2004). (Middle and Bottom) Conceptual model for the hierarchical sedimentary architecture found in channel belt deposits (see also Table 1). The compound bar deposits at level III result from the processes of unit-bar accretion and channel migration. Within unit bar deposits (level II), sets of open-framework gravel (level I) have highest permeability. As channels are abandoned, they are filled with lower-permeability sediment. Major channel fills (level III) and smaller cross-bar channel fills (level II) are lower-permeability baffles within the deposit. From Ramanathan et al. (2010).

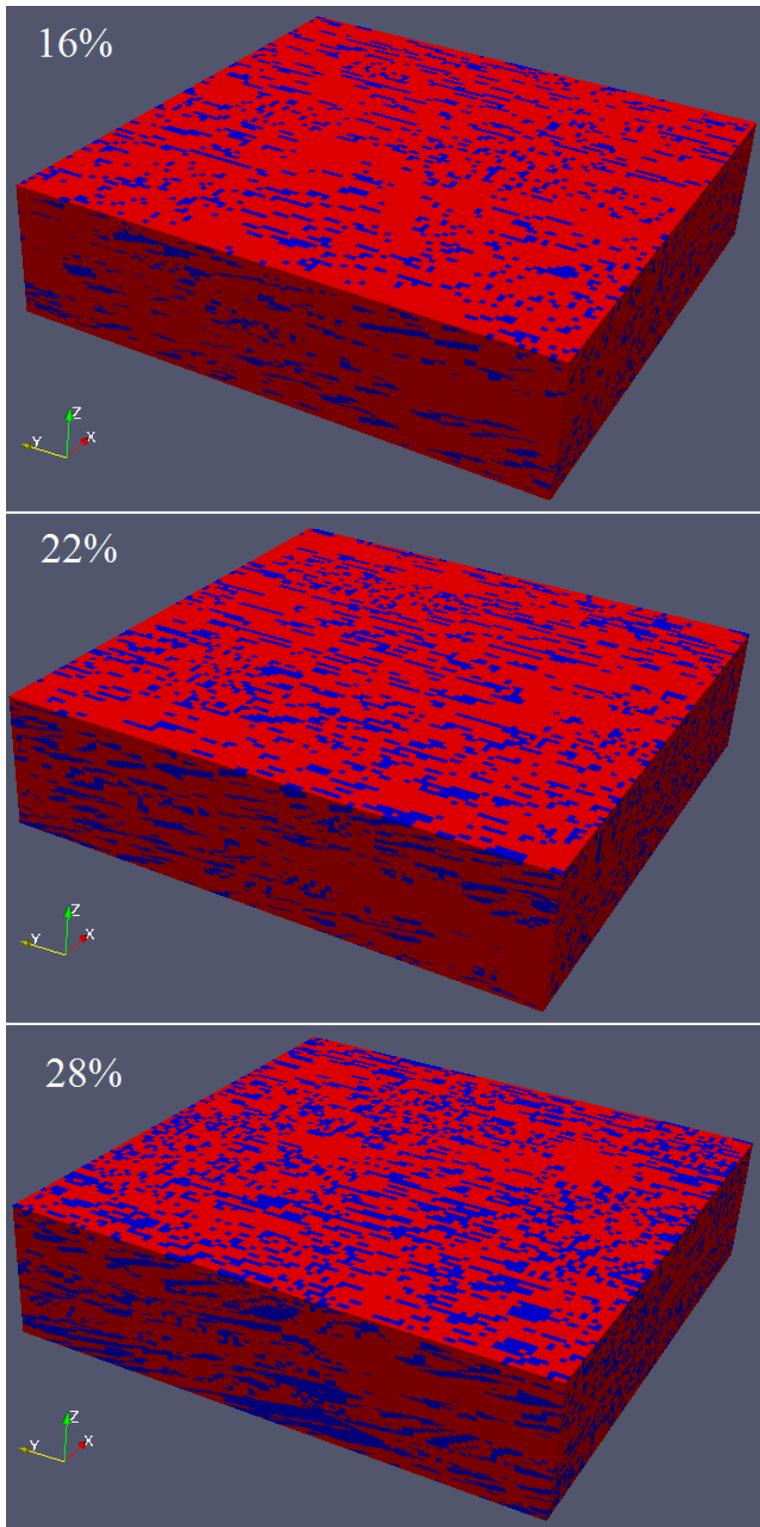

Figure 2. Distribution of sand (red) and OFG (blue) in realizations 6 (16% of OFG; top panel), 4 (22% of OFG; panel in the middle) and 1 (28% of OFG; bottom panel). Paleoflow during deposition is from right to left.

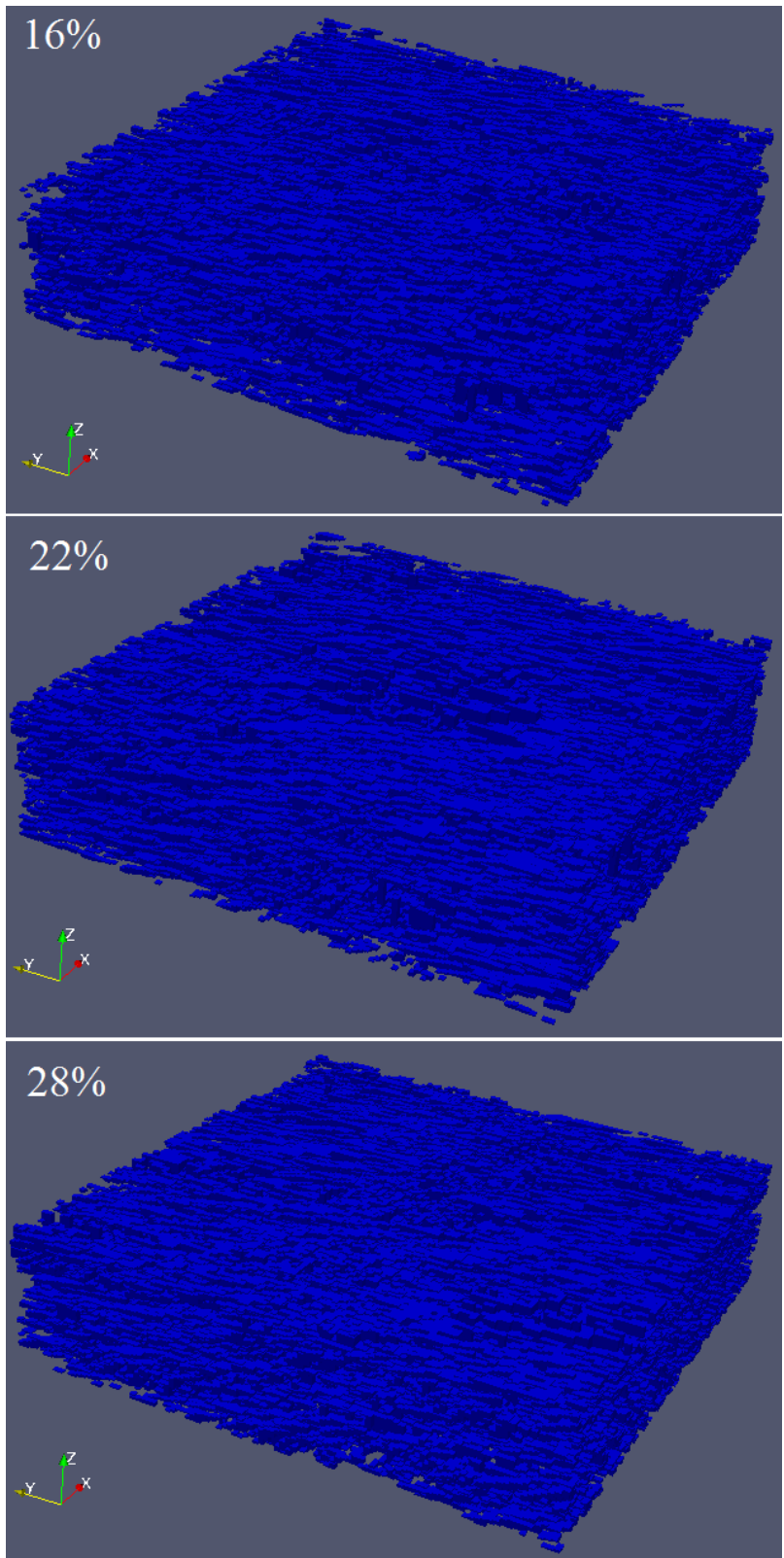

Figure 3A. Distribution of OFG in realizations 6 (16% of OFG; top panel), 4 (22% of OFG; panel in the middle) and 1 (28% of OFG; bottom panel)

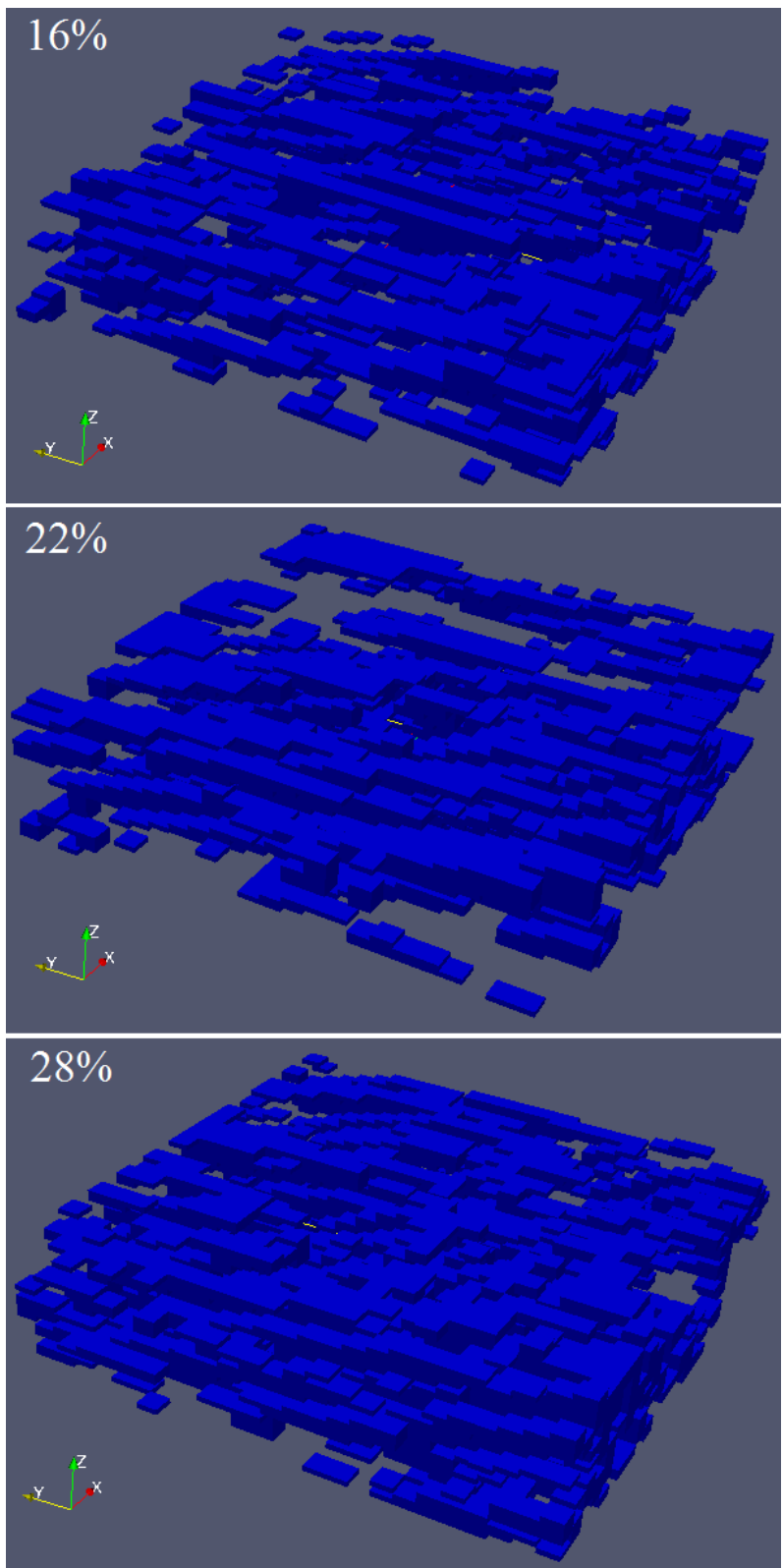

Figure 3B. Distribution of OFG in realizations 6 (16% of OFG; top panel), 4 (22% of OFG; panel in the middle) and 1 (28% of OFG; bottom panel) in reservoir with size (25:25:25)

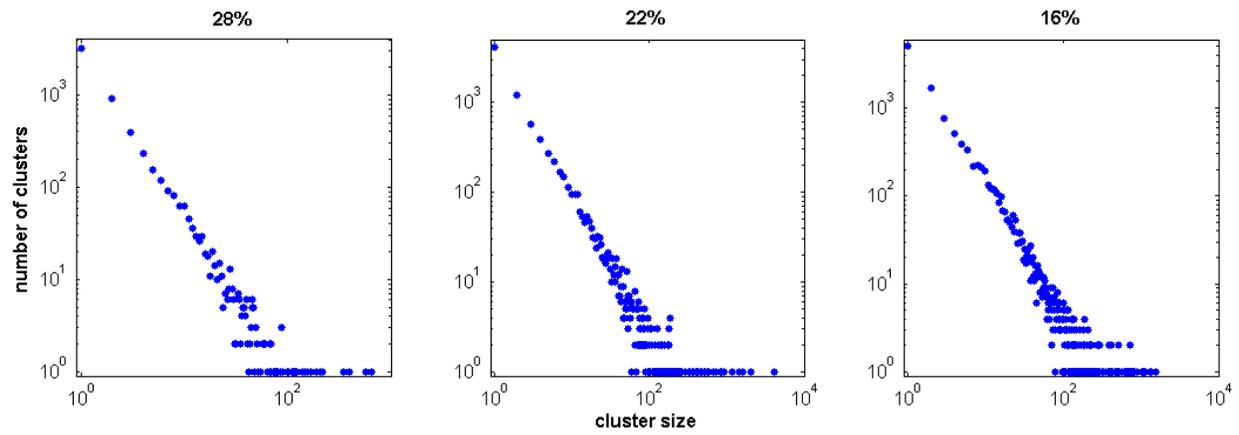

Figure 4. Distribution of OFG clusters by size (spanning clusters are not included).

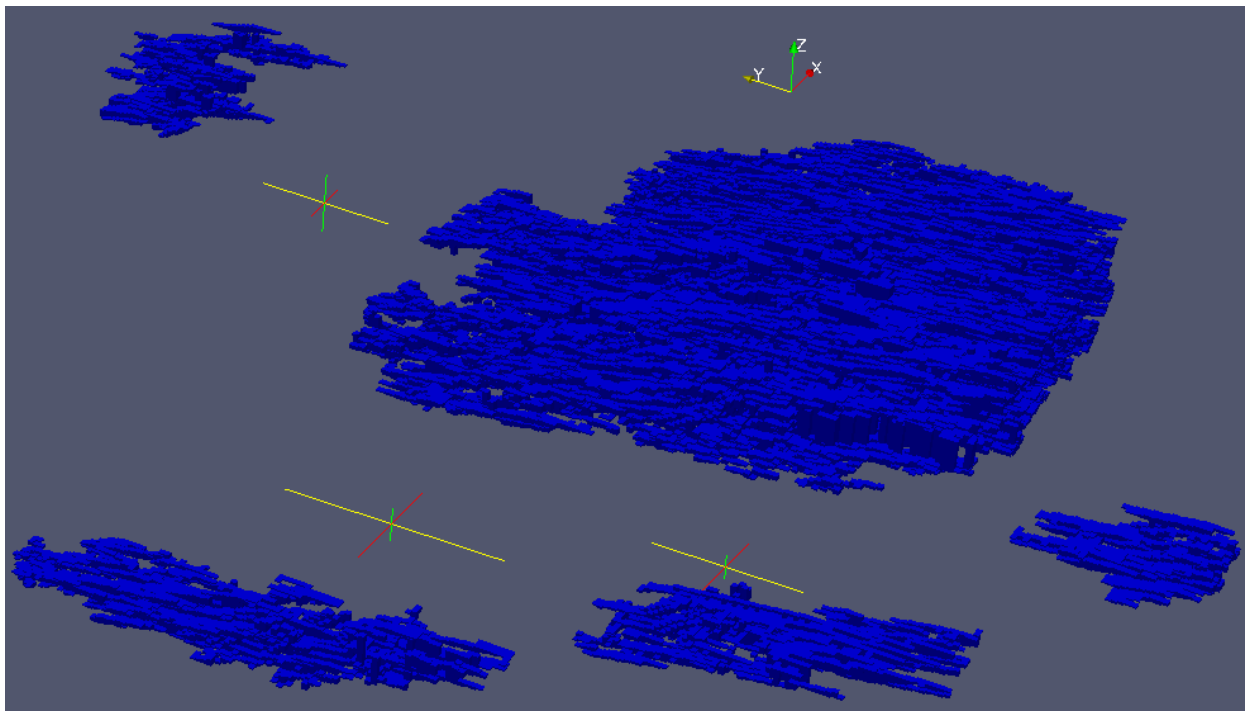

Figure 5. Distribution of OFG in realization 4 (22% of OFG). The 5 largest clusters are depicted after extracting them from the volume (each cluster is plotted in a separate coordinate system. The largest cluster in the middle is a spanning or infinite cluster.

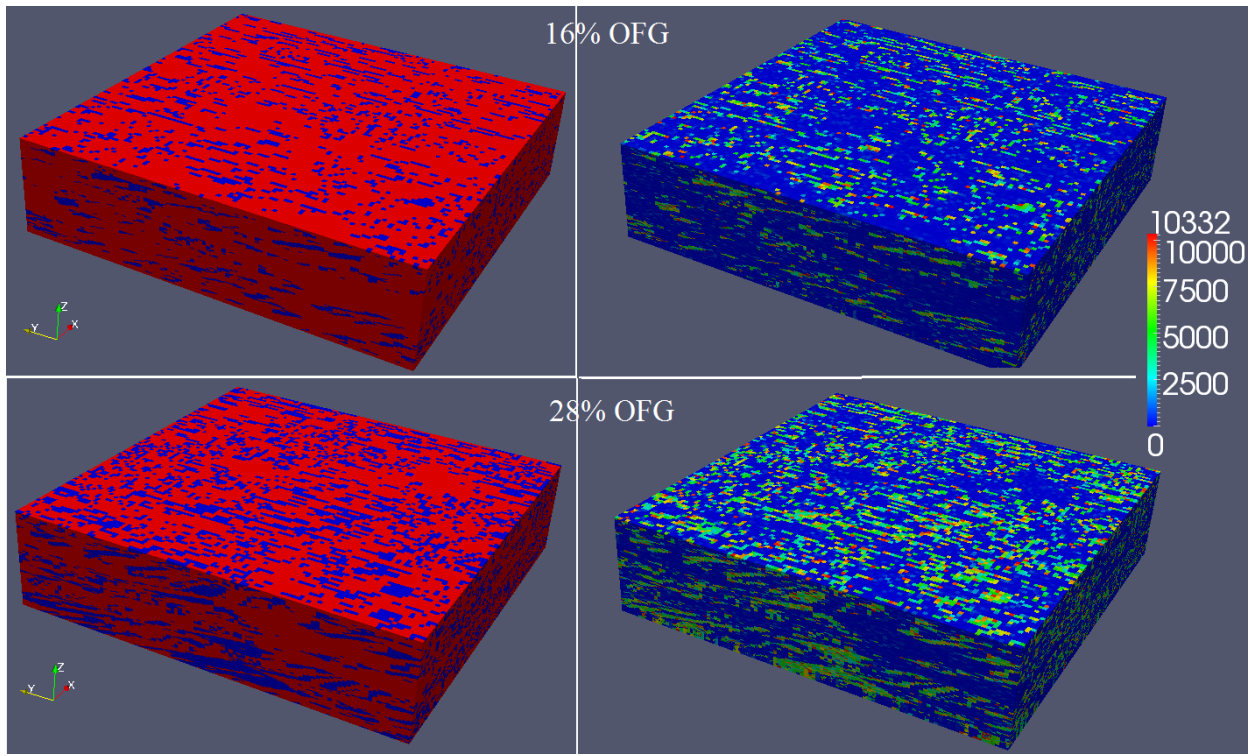

Figure 6. Image of reservoir with 16% (top) and 28% (bottom) of OFG. Panels on the left show distribution of sand (red) and OFG (blue) and panels on the right show the corresponding permeability map.

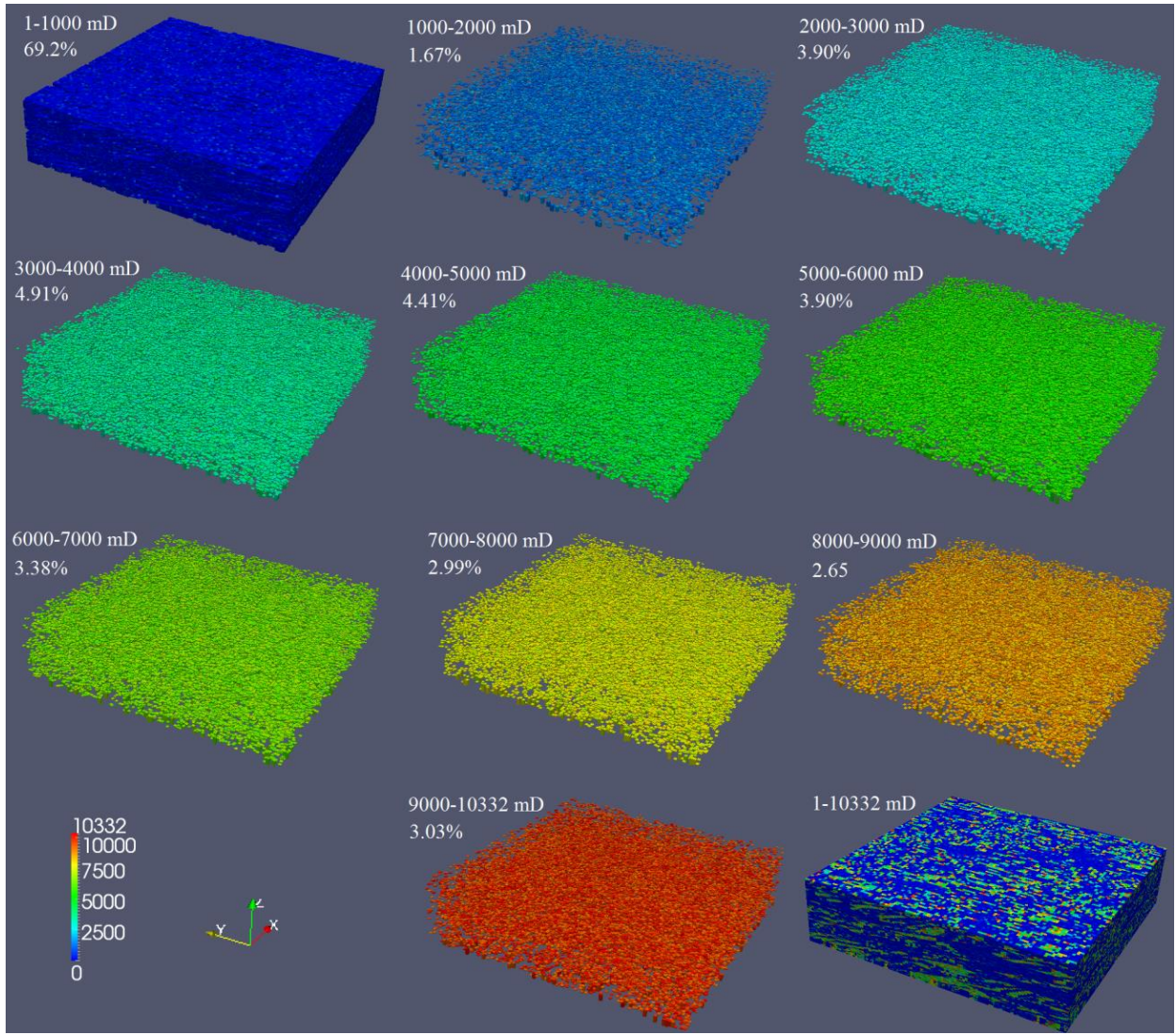

Figure 7. Images of permeability map for the realization of reservoir with 28% of OFG. Each image shows all cells with permeability in the range indicated at the top of the image. The percentage of cells with permeability in specified range from the total cell number is also shown.

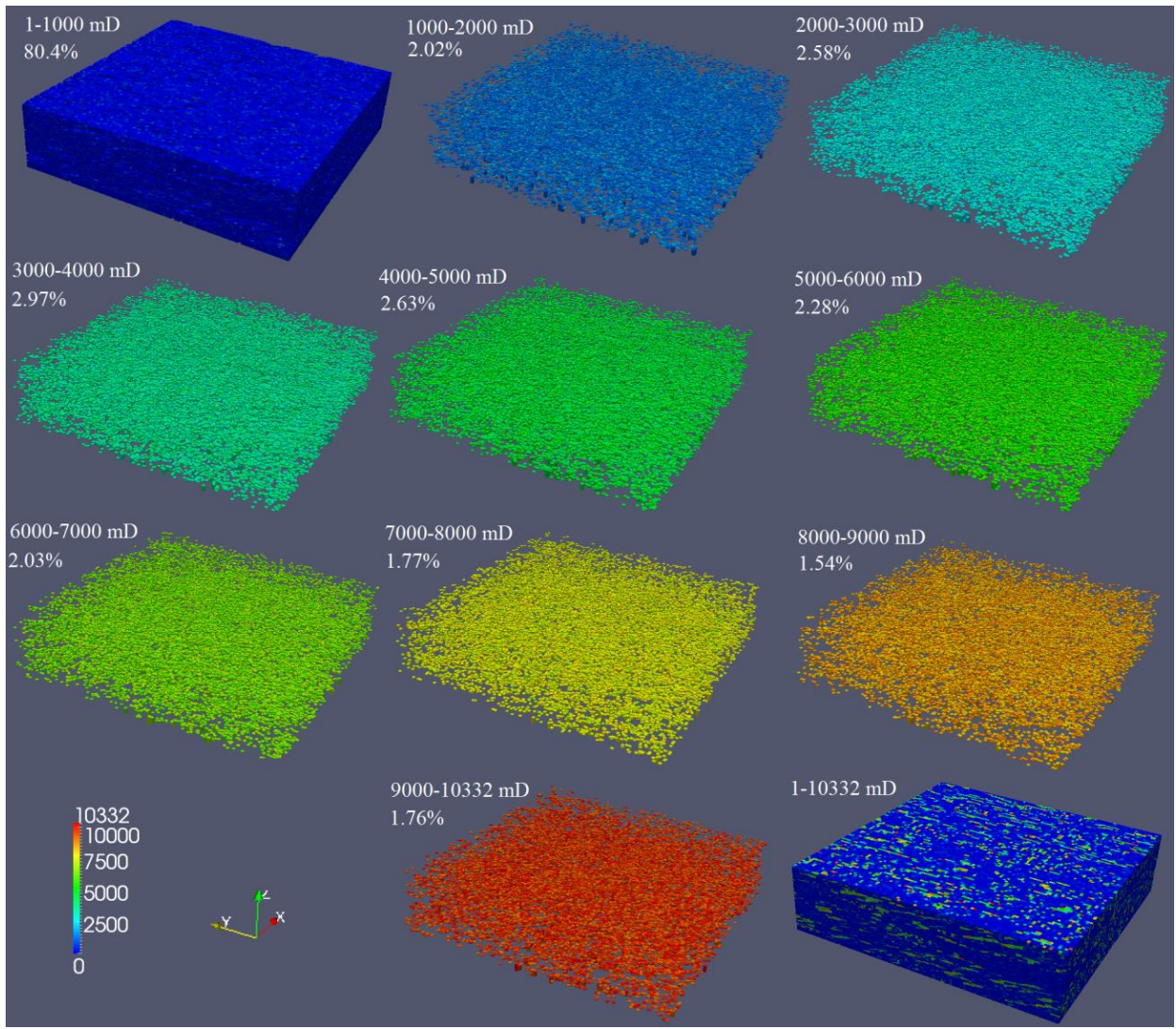

Figure 8. The same as Figure 7 for the realization of reservoir with 16% of OFG.

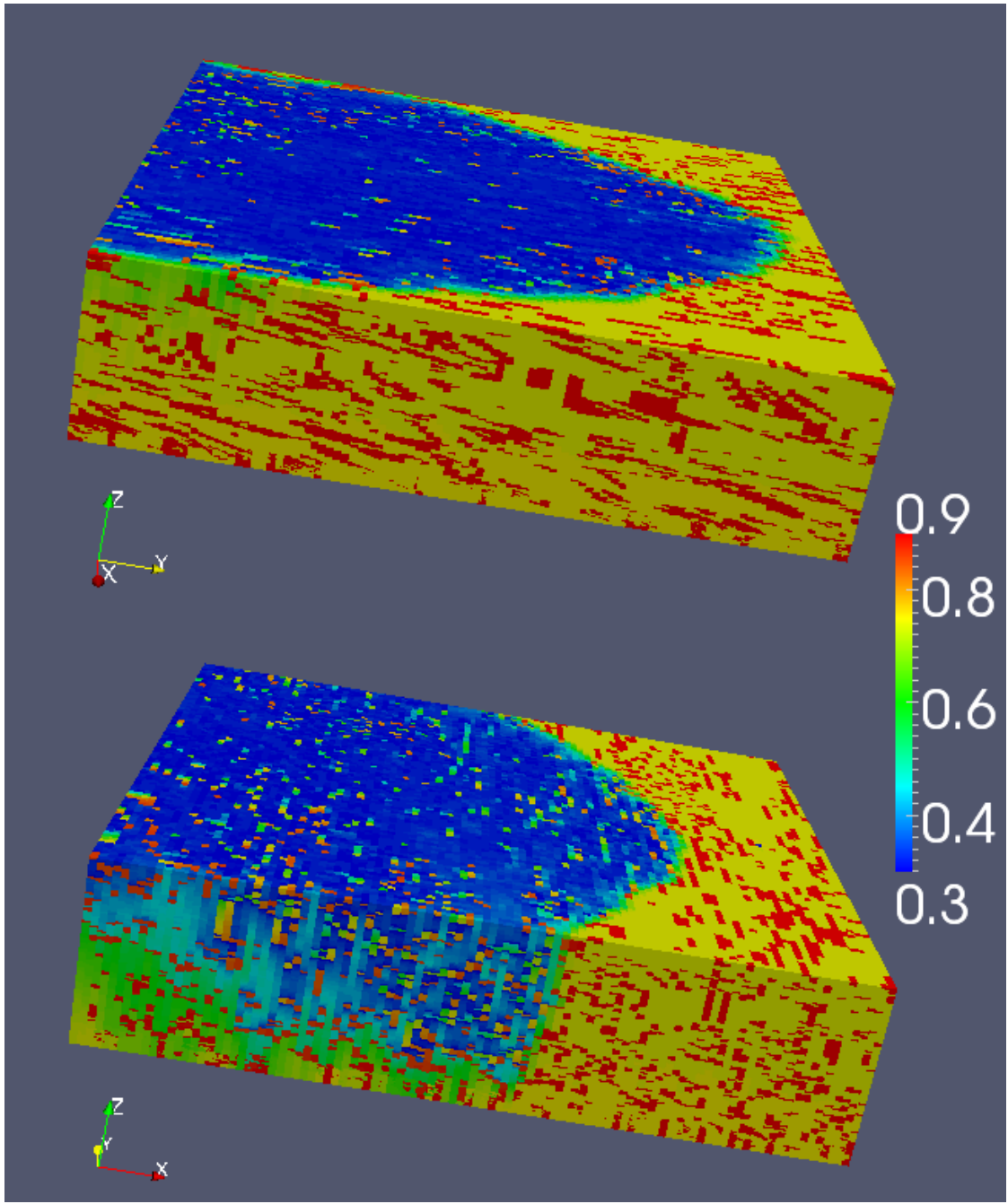

Figure 9. Oil saturation distribution in reservoir realization 4 after 200 days of water injection. Pressure difference between wells is 100 psi. Paleoflow was left to right in the top panel and front to back in the lower panel. Pressure gradient is parallel (top panel) and normal (bottom panel) to paleoflow direction.

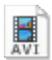
Figure 10.avi

Figure 10. Animated oil saturation distribution in reservoir realization 4 from the beginning of water injection up to 500 days. Pressure difference between wells is 100 psi, pressure gradient is parallel (a) and normal (b) to paleoflow direction.

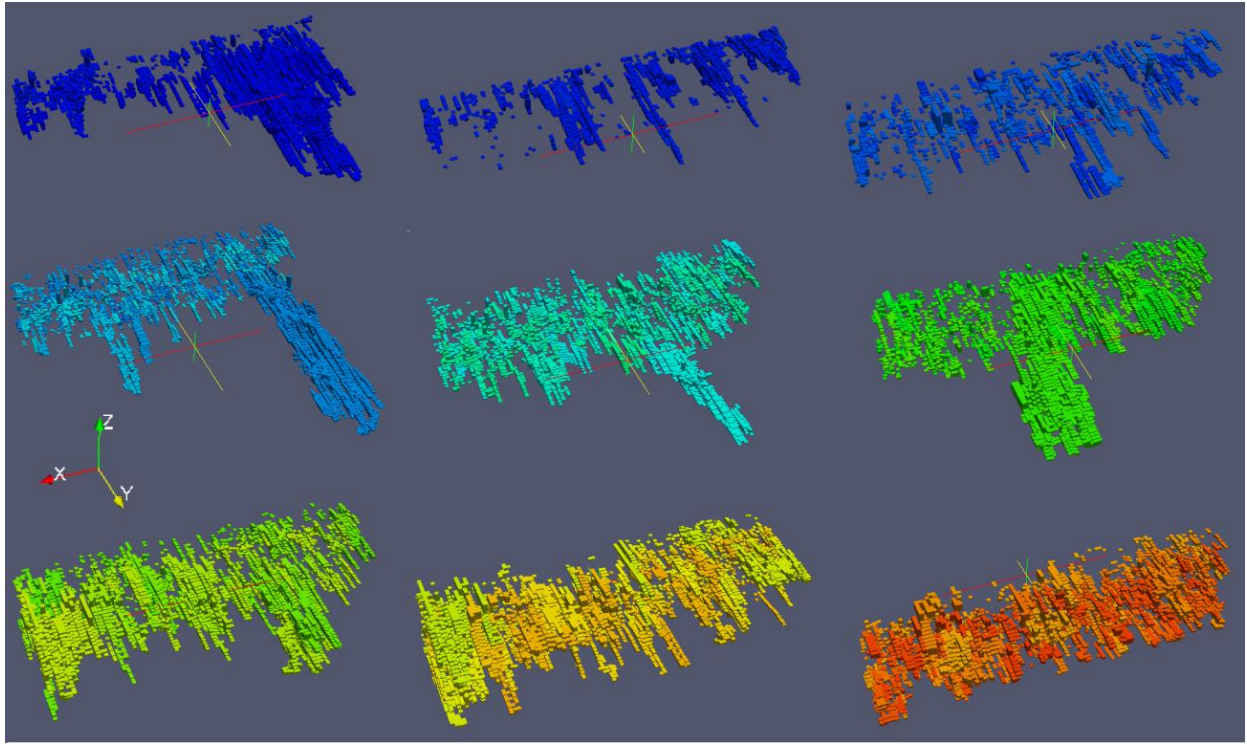

Figure 11. View of all (but spanning) OFG clusters (total number of clusters is 8562) from realization 4. Most of the clusters are elongate in paleoflow ($y$) direction.

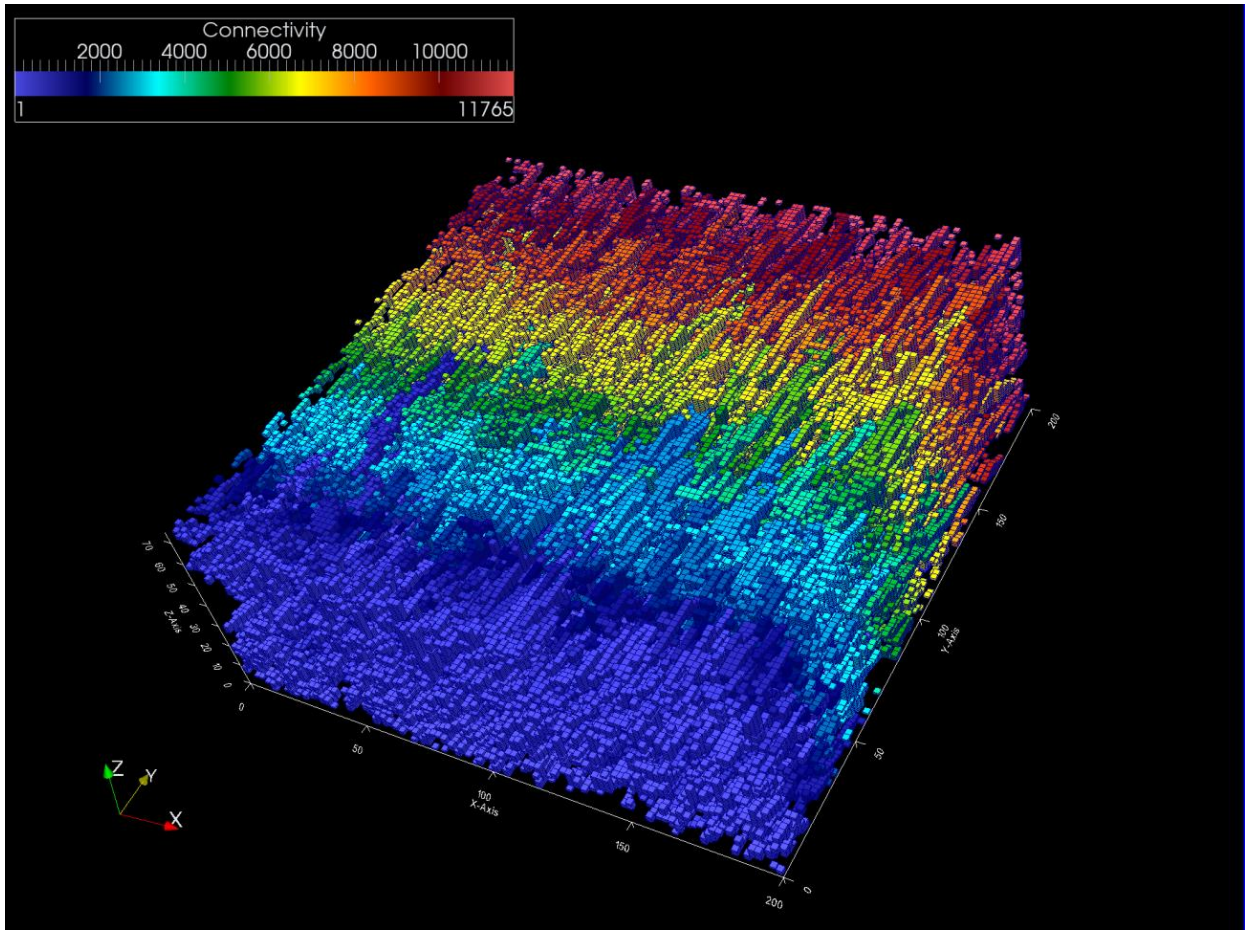

Figure 12a. View of all 11765 OFG clusters from realization 6.

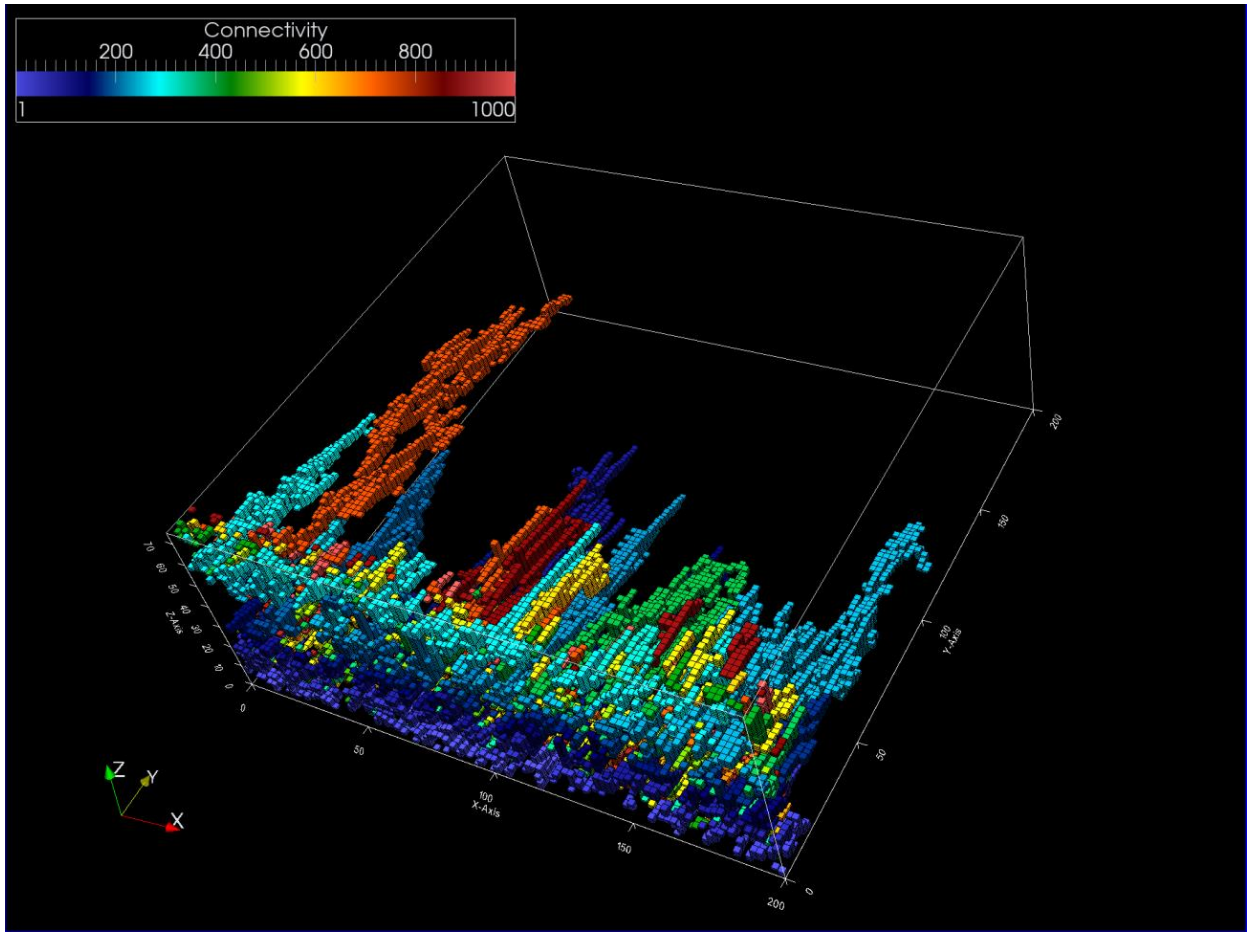

Figure 12b. View of first 1000 OFG clusters (total number of clusters is 11765) from realization 6.

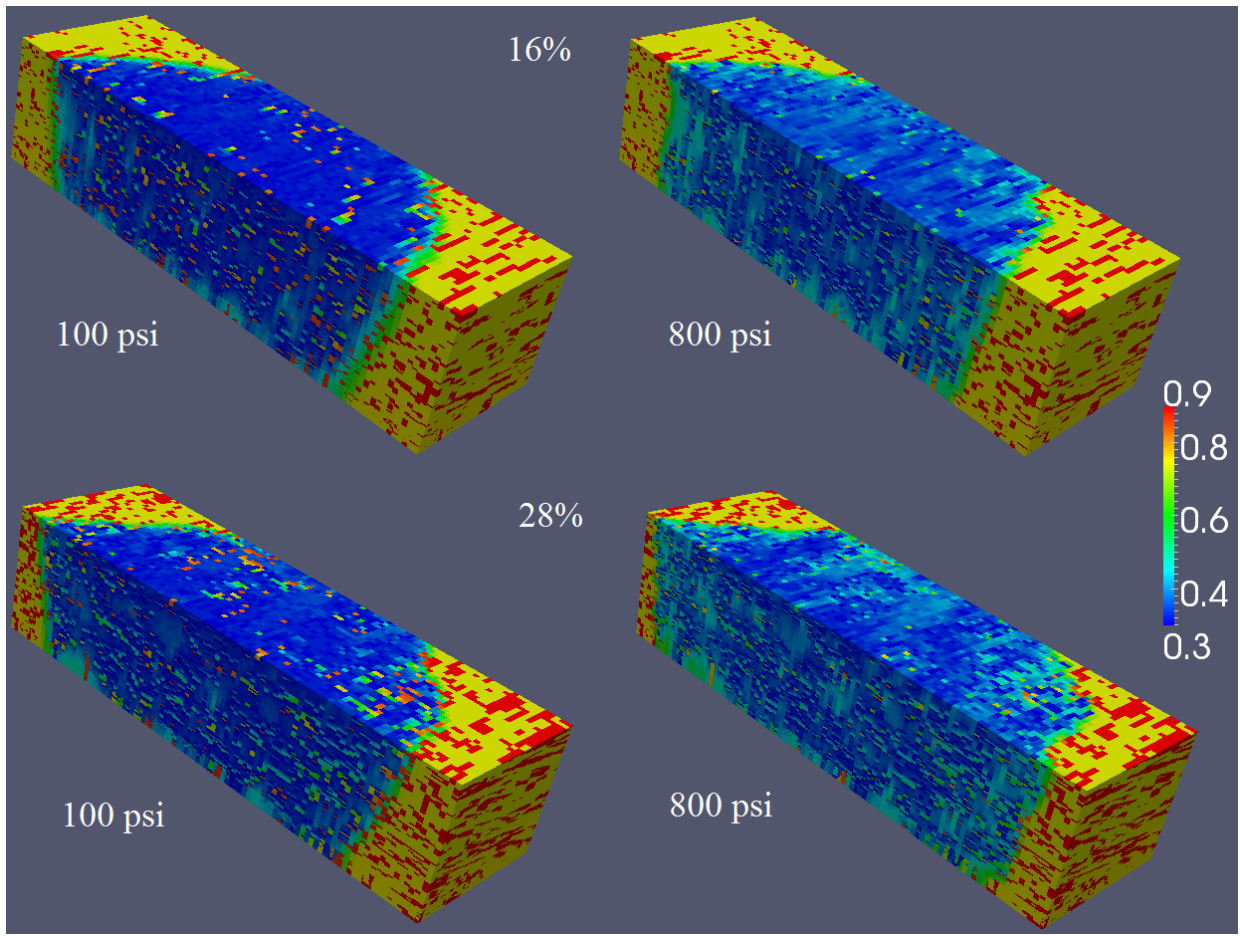

Figure 13. Oil saturation distribution for the slices of reservoir 0:200 x 100:160 x 1:5 m for the realization 1 (bottom panels) and 6 (top panels). Pressure difference between wells is 100 psi (left panels) and 800 psi (right panels).

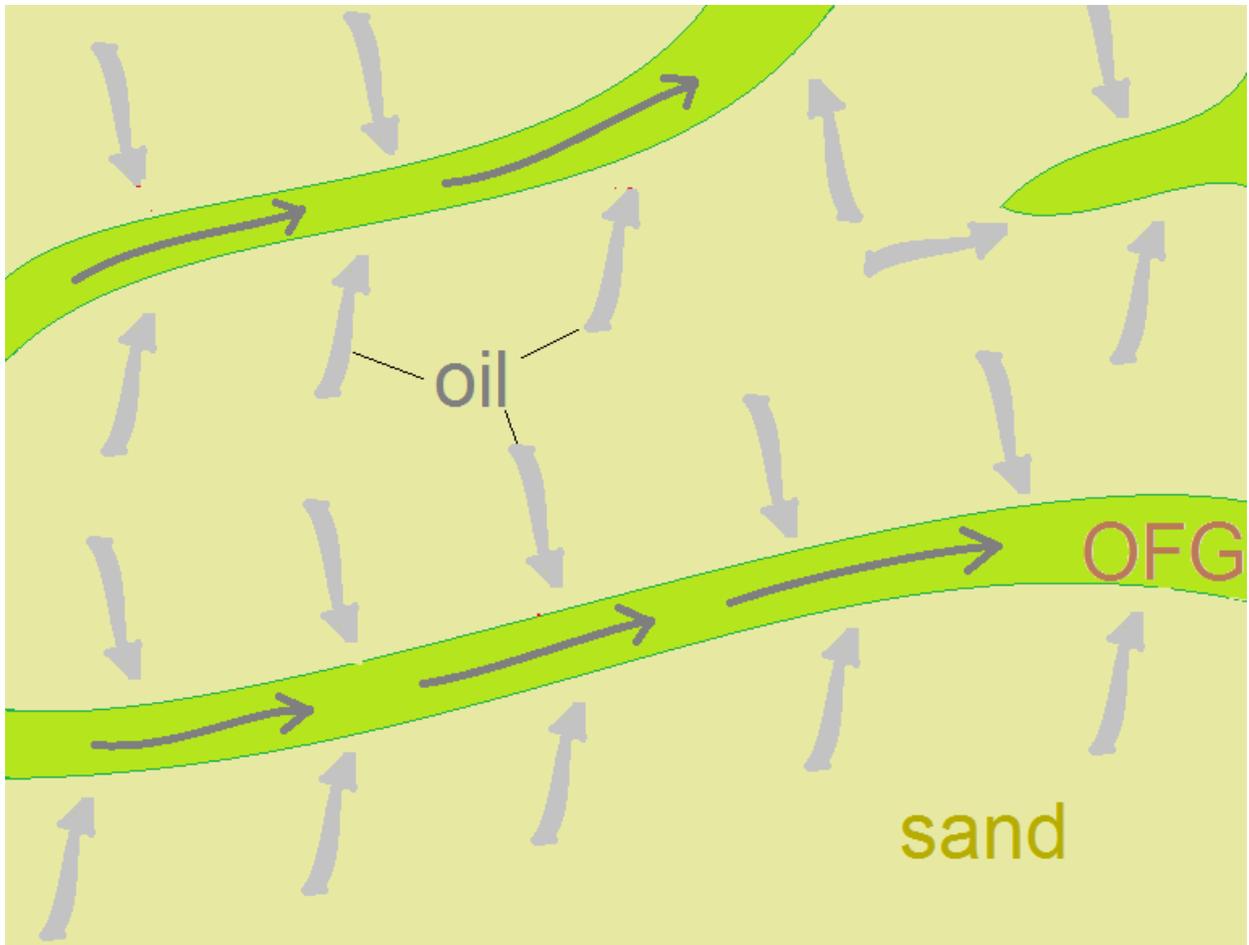

Figure 14. Image of oil path in fluvial-type reservoirs during waterflooding.

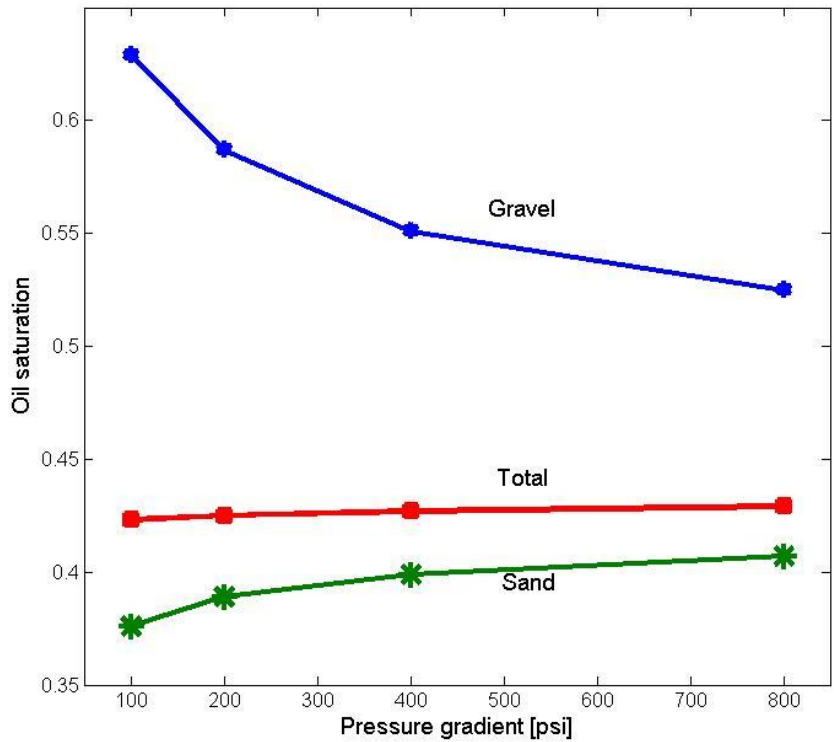

**Figure 15.** Oil saturation averaged over open gravel cells (blue), sand cells (green) and all cells (red) after injection of 1 water volume (in units of pore space volume minis volume of connate water and volume of irreducible oil) as function of pressure difference for Sample #5. From Gershenzon et al, 2014.

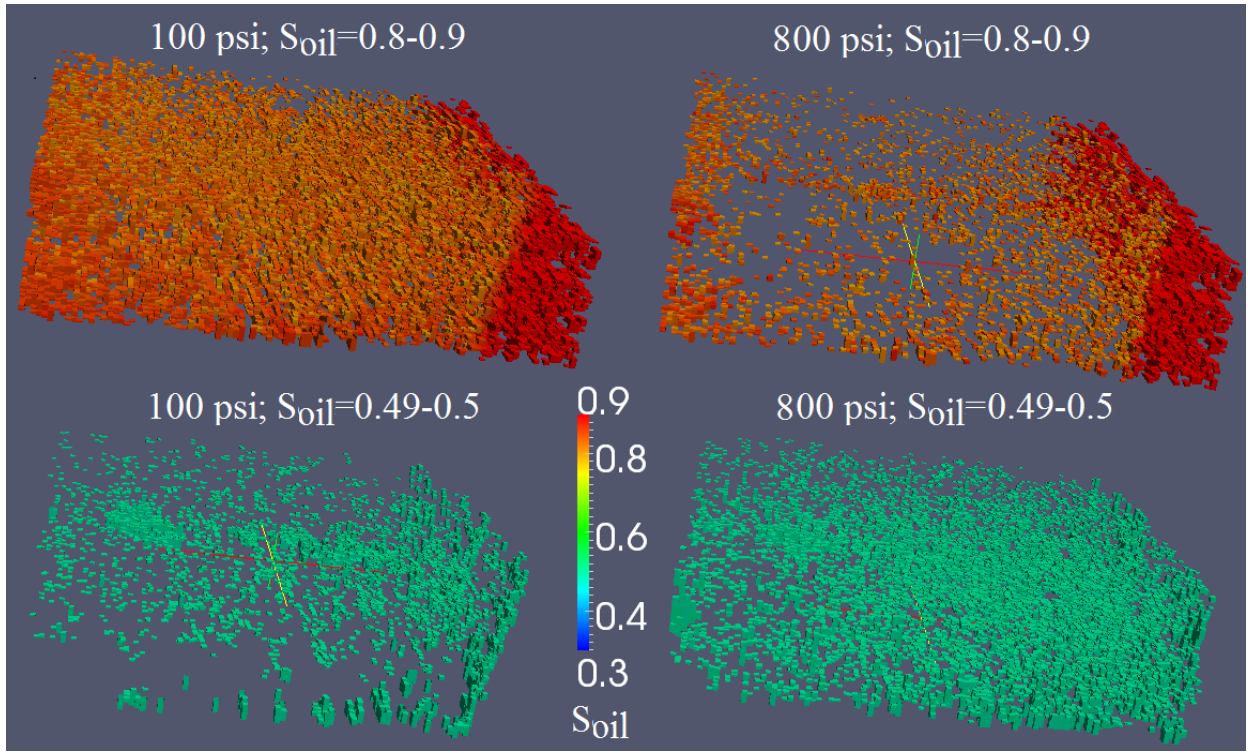

Figure 16. Cells with oil saturation in the range indicated at the top of each panels for the reservoir realization 6 (16% of OFG) after waterflooding with pressure difference between injection and production wells 100 psi (panels at the left) and 800 psi (panels at the right). The choice of intervals of oil saturation is arbitrary, i.e. 0.49-0.5 for sand and 0.8-0.9 for OFG. The intervals were chosen to make effect more visible.

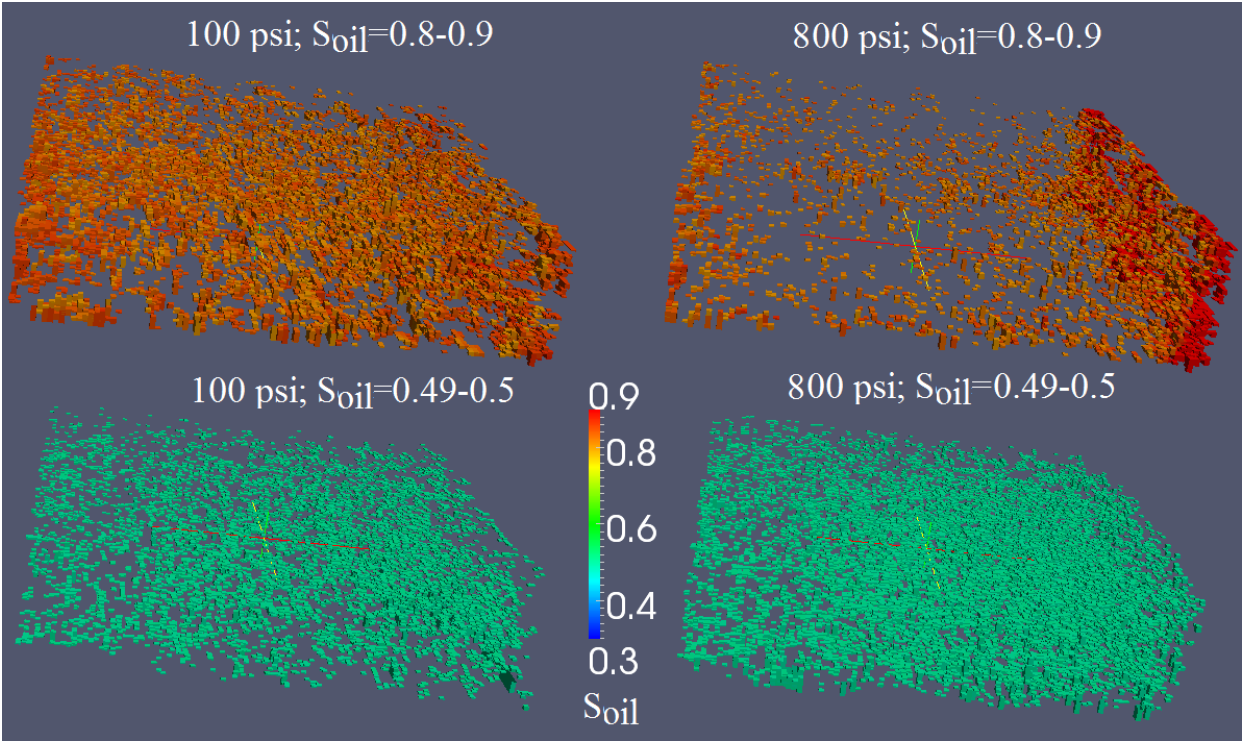

Figure 17. The same as at Figure 15 for the reservoir realization 1 (28% of OFG).